\documentclass[fleqn,10pt]{wlscirep} \title{Scaling crossover in thin-film drag dynamics of fluid drops in the Hele-Shaw cell} \author[]{Misato Yahashi} \author[]{Natsuki Kimoto} \author[1,*]{Ko Okumura} \affil[]{Department of Physics and Soft Matter Center, Ochanomizu University, 2--1--1, Otsuka, Bunkyo-ku, Tokyo 112-8610, Japan} \affil[*]{corresponding author: okumura@phys.ocha.ac.jp} 

\begin{abstract}
We study both experimentally and theoretically the descending motion due to
gravity of a fluid drop surrounded by another immiscible fluid in a confined
space between two parallel plates, i.e., in the Hele-Shaw cell. As a result,
we show a new scaling regime of a nonlinear drag friction in viscous liquid
that replaces the well-known Stokes' drag friction through a clear collapse of
experimental data thanks to the scaling law. In the novel regime, the
dissipation in the liquid thin film formed between the drop and cell walls
governs the dynamics. The crossover of this scaling regime to another scaling
regime in which the dissipation inside the droplet is dominant is clearly
demonstrated and a phase diagram separating these scaling regimes is presented.

\end{abstract}
\begin{document}

\flushbottom\maketitle
\thispagestyle{empty}

\section*{Introduction}

Dynamics of liquid drops is familiar in daily life: we observe rain drops
rolling on a new umbrella, honey dripping off from a spoon, and oil droplets
floating on the surface of vegetable soup and so on. Such everyday phenomena
are in fact important not only in physical sciences
\cite{RichardClanetQuere2002,DoshiCohenZhangSiegelHowellBasaranNagel2003,CouderProtiereFortBoudaoud2005,RistenpartBirdBelmonteDollarStone2009,PhysRevLett.79.1265,BirdDeCourbinStone2010,EtienneWedge2014}
but also in a variety of practical issues such as ink-jet printing
\cite{Calvert2001}, microfluidics manipulations
\cite{SquiresQuake2005,MathileTabeling2014}, and emulsification, formation of
spray and foams \cite{DynamicsDroplets,PhysicsFoams,Sylvie}. From such
phenomena familiar to everybody, researchers have successfully extracted a
number of scaling laws representing the essential physics \cite{CapilaryText},
which include scaling laws associated with the lifetime of a bubble in viscous
liquid \cite{DebregeasGennesBrochard-Wyart1998,EriOkumura2007} and contact
dynamics of a drop to another drop
\cite{AartsLekkerkerkerGuoWegdamBonn2005,YokotaPNAS2011} or to a solid plate
\cite{BirdMandreStone2008,DavidCoalPRE}. Here, we report on a crossover of two
scaling regimes experimentally revealed for viscous friction acting on a fluid
drop in a confined space. In particular, we study the descending motion (due
to gravity) of an oil droplet surrounded by another immiscible oil in a
Hele-Shaw cell. The friction law thus revealed is nonlinear and replaces the
well-known Stokes' law in the Hele-Shaw cell geometry.

A closely related topic of the rising bubble in a Hele-Shaw cell is
theoretically discussed by Taylor and Saffman in a pioneering paper
\cite{TAYLORSAFFMAN1959} in 1958 (earlier than the Bretherton's paper on
bubbles in tubes \cite{Bretherton,Clanet2004}). The solution of Taylor and
Saffman was further discussed by Tanveer \cite{Tanveer1986}. There are many
other theoretical works on fluid drops in the Hele-Shaw cell geometry, notably
in the context of the topological transition associated with droplet breakup
\cite{Eggers1997,ConstantinDupontGoldsteinKadanoffShelleyZhou1993,GoldsteinPesciShelley1995,Howell1999}%
. \ As for experimental studies, a number of researchers have investigated the
rising motion of a bubble in a Hele-Shaw cell
\cite{Maxworthy1986,Kopf-SillHomsy1988,MaruvadaPark1996}. However, unlike the
present study, systematic and quantitative studies in a constant velocity
regime have mostly concerned with the case in which there is a forced flow in
the outer fluid phase and most of the studies have been performed with the
cell strongly inclined nearly to a horizontal position (one of a few examples
of the case with the cell set in the upright position but with external flow
\cite{HeleShawPetroleum2010} demonstrates relevance of the present work to
important problems in petroleum industry, such as the suction of crude oil
from the well).\ 

One of the features of the present study compared with most of previous ones
on the dynamics of fluid drops in a Hele-Shaw cell is that in the present case
the existence of a thin liquid film surrounding a fluid drop plays a crucial
role: In many previous works, the existence of such thin films is not
considered. In this respect, the present problem is closely related to the
dynamics governed by thin film dissipation such as the imbibition of textured
surfaces
\cite{StoneNM2007,IshinoReyssatEPL2007,ObaraPRER2012,TaniPlosOne2014,TaniSR2015,DominicVellaImbibition2016}%
. In this sense, our problem is quasi two-dimensional, although the geometry
of the Hele-Shaw cell is often associated with a purely two-dimensional problem.

\section*{Experiment}

\begin{figure}[ptb]
\begin{center}
\includegraphics[width=0.7\linewidth]{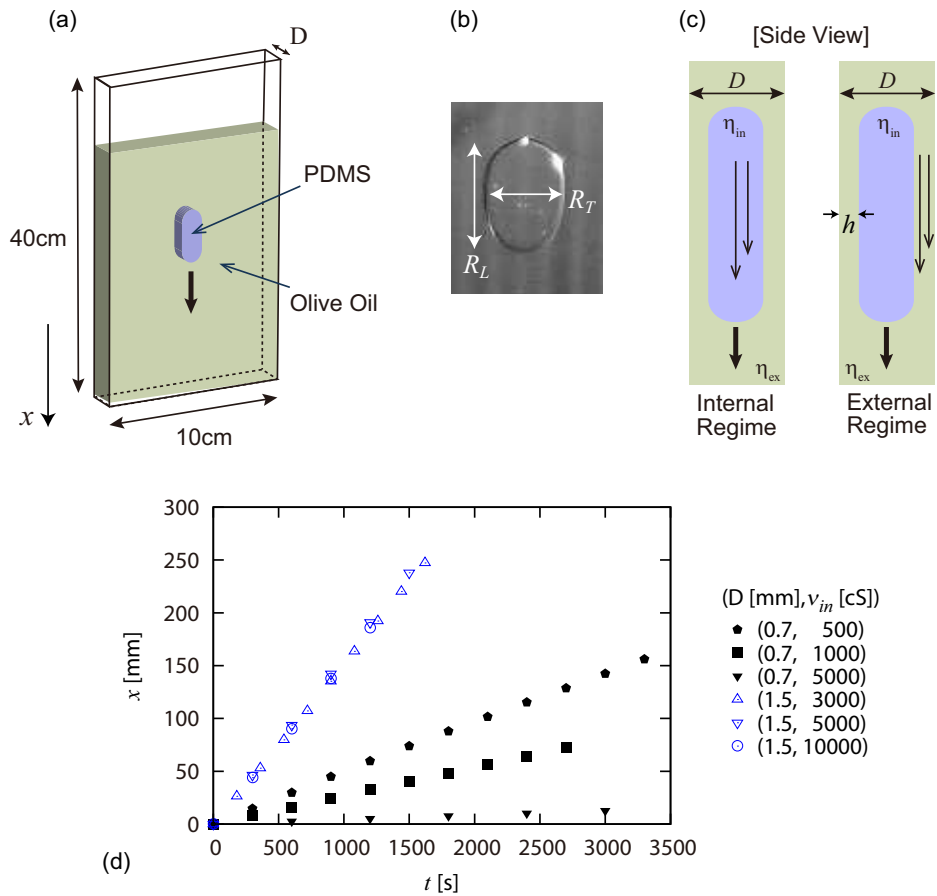}
\end{center}
\caption{(a) Experimental setup. Gravity is acting in the $x$-direction. (b)
Front view of a PDMS drop of kinematic viscosity $\nu_{in}=1000$ cS going down
in olive oil in a Hele-Shaw cell of thickness $D=2$ mm. (c) Magnified side
views of droplets with the velocity gradient in the internal end external
regimes. (d) Position of the PDMS drop $x$ as a function of elapsed time $t$.
}%
\label{Fig1}%
\end{figure}

We fabricated a Hele-Shaw cell of thickness $D$
\cite{EriOkumura2007,EriOkumura2010,YokotaPNAS2011} and filled the cell with
olive oil (150-00276, Wako; kinematic viscosity $\nu_{ex}=60$ cS and density
$\rho_{ex}=910$ kg/m$^{3}$). This oil plays a role of an external surrounding
liquid for a drop of poly(dimethylsiloxane) (PDMS) to be inserted at the top
of the cell using a syringe (SS-01T, Termo). We observe the inserted drop
going down in the cell, as illustrated in Fig. \ref{Fig1}(a), because of the
density difference $\Delta\rho=\rho_{in}-\rho_{ex}>0$. The drop density
$\rho_{in}$ depends on its kinematic viscosity $\nu_{in}$ only slightly (see
the details for Methods). The drop size is characterized by the cell thickness
$D$ and the width $R_{T}$, i.e., the size in the direction transverse to that
of gravity (see Fig. \ref{Fig1}(b)), which is slightly smaller than the size
in the longitudinal direction, $R_{L}$. As shown in Fig. \ref{Fig1}(c), a thin
film of olive oil exists between a cell plate and the surface of the drop. We
can think of two limiting cases for the distribution of liquid flow: (1)
Internal Regime: The velocity gradient is predominantly created in the
internal side of the droplet as in the left illustration. (2) External Regime:
The gradient is predominantly exists in the external side of the droplet as in
the right.

The width and height of the cell are 10 cm and 40 cm, respectively, and are
much larger than the drop size to remove any finite size effects in the
direction of width and height. The cell is made of acrylic plates of thickness
5 mm, to avoid thinning deformation of the cell due to the effect of capillary
adhesion \cite{CapilaryText}.

We took snapshots of the descending drop at a regular time interval using a
digital camera (Lumix DMC-G3, Panasonic) and a camera controller (PS1,
Etsumi). The obtained data were analyzed with the software, Image J, to obtain
the position as a function of time to determine the descending velocity of the
drop. Some examples are shown in Fig. \ref{Fig1}(d). This plot show the
following facts. (1) The descending motion can be characterized by a
well-defined constant velocity (to guarantee a long stationary regime, the
cell height is made significantly larger (40 cm) than the drop size; because
of a small density difference, the constant-velocity regime starts after a
long transient regime). (2) The descending velocity is dependent on the
kinematic viscosity of the internal liquid of the drop $\nu_{in}$ for the
thinner cell ($D=0.7$ mm) as predicted in the previous study
\cite{EriSoftMat2011}, which is not the case for the thicker cell ($D=1.5$
mm); These examples clearly demonstrate the existence of a novel scaling
regime different from the one discussed in the previous study
\cite{EriSoftMat2011}.

In the present study, the dependence of the descending velocity on the drop
size is negligible. In the previous study \cite{EriSoftMat2011}, it was found
that the descending speed of drops is dependent on $R_{T}$ for $R_{T}/D<10$ if
a glycerol drop goes down in PDMS oil. However, in the present combination
(i.e., a PDMS drop going down in olive oil), we do not observe a significant
dependence on $R_{T}$ in our data even for fairly small drops, whereas $R_{T}$
is in the range $1.31<R_{T}/D<15.8$ in the present study (the size dependence
in the previous study may be caused by the polarity of the glycerol aqueous
solution: We expect that if the liquid is polar, the drop may subject to
electrostatic (attractive) force from the acrylic cell plates and this effect
tends to make the drop less mobile).\ The data analysis below neglects any
possible small dependences of the velocity on the drop size.

\section*{Theory}

At the level of scaling laws, the characteristic energy scales are given as
follows. The gravitational energy gain for the descending drop per unit time
is expressed as%

\begin{equation}
\mathcal{\dot{E}}=\alpha\Delta\rho gR_{T}R_{L}DV, \label{eq3}%
\end{equation}
where $\alpha$ is a numerical coefficient. The viscous dissipation per unit
time in the internal regime discussed above (Fig. \ref{Fig1}(c) left) is
written as%

\begin{equation}
T\dot{S}_{in}=(\alpha/k_{in})\eta_{in}(V/D)^{2}R_{T}R_{L}D, \label{eq1}%
\end{equation}
where $\alpha/k_{in}$ is a numerical coefficient. Strictly speaking, because
of the existence of the thin film of thickness $h$ (Fig. \ref{Fig1}(c)), the
velocity gradient $V/D$ in the above expression should be replaced with
$V/(D-2h)$, which is not essential, however, because the relation $D\gg h$ is
well satisfied in the present study (see the next section). The viscous
dissipation per unit time in the external regime discussed above (Fig.
\ref{Fig1}(c) right) is given as%

\begin{equation}
T\dot{S}_{ex}=(\alpha/k_{ex})\eta_{ex}(V/h)^{2}R_{T}R_{L}h, \label{eq2}%
\end{equation}
where $\alpha/k_{ex}$ is a numerical coefficient.

In the internal regime the velocity is given by the balance between $T\dot
{S}_{in}$ and $\mathcal{\dot{E}}$,%

\begin{equation}
V_{in}=k_{in}\Delta\rho gD^{2}/\eta_{in}, \label{eq4}%
\end{equation}
whereas in the external regime the velocity is given by the balance between
$T\dot{S}_{ex}$ and $\mathcal{\dot{E}}$,%

\begin{equation}
V_{ex}=k_{ex}\Delta\rho gDh/\eta_{ex}. \label{eq5}%
\end{equation}

The thickness of the thin film formed between the drop and cell plates may be
given by the law of Landau, Levich and Derjaguin (LLD),%

\begin{equation}
h=k_{LL}Ca^{2/3}\kappa^{-1}, \label{eq6}%
\end{equation}
where the numerical coefficient is of the order of unity \cite{CapilaryText}
($k_{LL}=0.94$, in the original papers \cite{LandauLevich,Derjaguin1943}).
Here the capillary length $\kappa^{-1}$ is defined as $\kappa^{-1}%
=\sqrt{\gamma/(\Delta\rho g)}$, which is smaller than the cell thickness $D$
(If otherwise, the length $\kappa^{-1}$ is replaced with $D$ with the
coefficient $k_{LL}=1.337$ \cite{ParkHomsyHeleShaw1984}). The capillary number
$Ca$ is defined as
\begin{equation}
Ca=\eta_{ex}V_{ex}/\gamma\label{eq7}%
\end{equation}

Removing $h$ from Eqs. (\ref{eq5}) and (\ref{eq6}), we obtain another
expression for the velocity in the external regime,%
\begin{equation}
\eta_{ex}V_{ex}/\gamma=(k_{1}D/\kappa^{-1})^{3} \label{eq8}%
\end{equation}
with $k_{1}=k_{ex}k_{LL}$. Removing $V_{ex}$ from Eqs. (\ref{eq5}) and
(\ref{eq6}), we obtain an expression for the thickness of the thin film,%
\begin{equation}
h/\kappa^{-1}=(k_{2}D/\kappa^{-1})^{2} \label{eq9}%
\end{equation}
with $k_{2}=k_{LL}^{3/2}k_{ex}=k_{LL}^{1/2}k_{1}$.

The condition for the internal regime is given by $T\dot{S}_{in}<T\dot{S}%
_{ex}$, which leads to the equation, $\eta_{ex}/\eta_{in}>k_{3}D/\kappa^{-1}$.
In other words, the phase boundary between the internal and external regimes
is given by
\begin{equation}
\eta_{ex}/\eta_{in}=k_{3}D/\kappa^{-1}\label{eq10}%
\end{equation}
with $k_{3}=k_{1}^{3}/k_{in}$. This means that the phase boundary between the
internal and external regime is a straight line with the slope $k_{3}$ on the
plot of $\eta_{ex}/\eta_{in}$ as a function of $D/\kappa^{-1}$.

\section*{Experiment and theory}

\begin{figure}[ptb]
\begin{center}
\includegraphics[width=\linewidth]{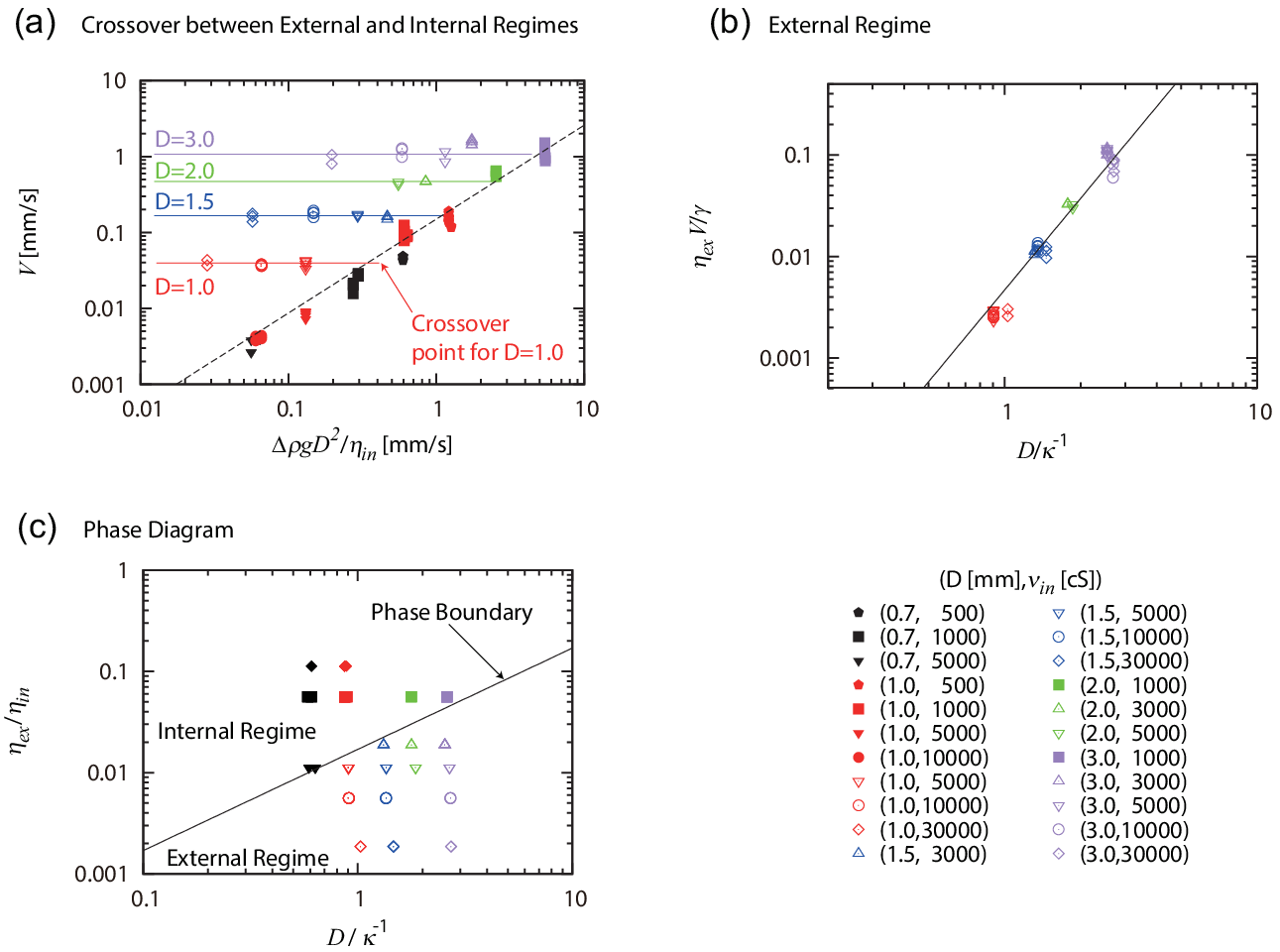}
\end{center}
\caption{(a) Plot of $V$ vs. $\Delta\rho gD^{2}/\eta_{i}$. The data in the
internal regime are on the dashed line, whereas the data in the external
regime are on horizontal lines for each cell thickness $D$. The crossover
between the two regimes would occur at the cross point of the dashed line and
each horizontal line. (b) Plot of $\eta_{ex}V/\gamma$ vs $D/\kappa^{-1}$,
confirming the external regime. (c) Plot of $\eta_{ex}/\eta_{in}$ vs
$D/\kappa^{-1}$, showing the phase diagram for the two scaling regimes.
Throughout (a)-(c), the data in the external and internal regimes are
represented by open and filled symbols, respectively. The color and shape of
the symbols are fixed for a given $D$ and a given $\nu_{in}$, respectively.}%
\label{Fig2}%
\end{figure}

The experimental data for the descending velocity of drops $V$ are plotted as
a function of $\Delta\rho gD^{2}/\eta_{in}$ in Fig. \ref{Fig2}(a). In view of
Eq. (\ref{eq4}), the data points in the internal regime would be on a straight
line of slope 1. This is almost true: there is a series of data well on the
dashed line of slope close to one. Naturally, there is a slight deviation from
the theory: the slope of the straight dashed line obtained by a numerical
fitting is in fact $1.24\pm0.06$, a value slightly larger than one, but the
coefficient corresponding $k_{in}$ is $0.150\pm0.015$, the order of magnitude
of which is consistent with the scaling arguments.

Some detailed remarks for the above arguments are as follows. (1) Even in the
previous study \cite{EriSoftMat2011} in which the internal scaling regime was
confirmed for the first time, the scaling regime described by Eq. (\ref{eq4})
was shown with some deviations, similarly to the present case (whereas another
scaling regime first established in the previous paper \cite{EriSoftMat2011}
is almost perfectly demonstrated). (2) We note here that the data represented
by the red filled circle and red filled inverse triangle are exceptional ones
and their seemingly strange behavior will be explained in Discussion. (3) We
have confirmed that even if we replace $D$ with $D-2h$ in the analysis (by
using the thickness $h$ estimated from Eq. (\ref{eq6})) when $D$ is used as a
length scale characterizing the viscous gradient  (i.e., when $D$ is used in
the expression $V/D$ in Eq. (2)), any visible differences are not introduced
into the plots given in Fig. \ref{Fig2} (This correction could be motivated by
considering the existence of thin films surrounding the drops as in Fig.
\ref{Fig1}(c) as mentioned above).

In Fig. \ref{Fig2}(b), it is shown that some of the data we obtained clearly
satisfy Eq. (\ref{eq8}), which describes the external regime. In Fig.
\ref{Fig2}(b), we collected the data points that are off the dashed line of
slope close to one in Fig. \ref{Fig2}(a) and that are thus ruled out from the
internal regime. The data thus selected and plotted in Fig. \ref{Fig2}(b) are
almost on the straight line of slope 3 in accordance with Eq. (\ref{eq8}). The
straight line is obtained by a numerical fitting with fixing the slope to 3.0;
as a result of this fitting, the coefficient is given as $k_{1}=0.167\pm
0.003$, the order of magnitude of which is consistent with the scaling arguments.

We confirm this scaling law in Eq. (\ref{eq8}) also in Fig. \ref{Fig2}(a). In
the light of Eq. (\ref{eq8}), the data in the external regime for a given $D$
should take almost the same values, because $\eta_{ex}$ and $\gamma$ are both
constant and $\kappa^{-1}$ is almost constant (note that $\Delta\rho$ is
almost constant) in the present study. In fact, in Fig. \ref{Fig2}(a), the
data points for a fixed $D$ that are off the dashed line, which data are shown
to be in the external regime in Fig. \ref{Fig2}(b), take almost a constant
value, that is, they are located almost on a horizontal line. This fact also
confirms that the data in question are independent of $\eta_{in}$, that is,
they are certainly not in the internal regime. Strictly speaking, the data
labeled as a given $D$ can have slightly different measured values of $D$ (see
Methods), which is the main reason the data for a "given" $D$ that are off the
dashed line in Fig. \ref{Fig2}(a) slightly deviate from the straight
horizontal line corresponding the $D$ value.

The scaling law in Eq. (\ref{eq8}) can be confirmed in Fig. \ref{Fig2}(a) in a
still another way. The open marks of the same shape, say diamond, but with
different colors (that are the data for a given $\nu_{in}$ but with different
$D$) are almost on a straight line of a slope close to one (This slope may
seem to be slightly larger than one, which may be because of the uncertainty
on the cell spacing $D$ as already mentioned\ in the last sentence of the
paragraph just above this one, or because the exponent 3 in Eq. (\ref{eq8})
may be in fact slightly larger than 3 in a more complete theory beyond the
present arguments at the level of scaling laws). For a such series of data,
the velocity $V$ scales with $D^{3}$ according to Eq. (\ref{eq8}), thus when
plotted as a function of $D^{2}$ as in Fig. \ref{Fig2}(a), the quantity
linearly scales with $D$, as \ reasonably well confirmed.

The phase diagram based on Eq. (\ref{eq10}) is shown in Fig. \ref{Fig2}(c), in
which\ we plot all the data (except for the special data mentioned above), to
demonstrate further consistency of the present arguments. As expected from Eq.
(\ref{eq10}), we can indeed draw a straight line of slope 1 on Fig.
\ref{Fig2}(c), which divides the internal and external regimes; Above the
straight line of slope 1 in Fig. \ref{Fig2}(c) lie the data in the internal
regime described by Eq. (\ref{eq4}), i.e., the data on the straight dashed
line in Fig. \ref{Fig2}(a); Below the straight line in Fig. \ref{Fig2}(c) lie
the data in the external regime described by Eq. (\ref{eq8}), i.e., the data
on the straight line in Fig. \ref{Fig2}(b). The coefficient $k_{3}$ of Eq.
(\ref{eq10}), i.e., the line dividing two regimes shown in Fig. \ref{Fig2}(c),
is $k_{3}=0.017$,\ the order of magnitude of which is consistent with the
scaling arguments in a profound\ sense: The numerical coefficient, $k_{in}$,
$k_{1}$, and $k_{3}$, are predicted to satisfy the relation $k_{3}=k_{1}%
^{3}/k_{in}$, and this relation is satisfied at a quantitative level in the
present analysis (0.017 vs $(0.167)^{3}/0.15\simeq0.031$). This quantitative
agreement is indeed quite satisfactory, if we consider slight deviations of
the data from the predicted theory. For example, the value $0.15$ used in the
estimation in the parentheses is not the value of $k_{in}$ itself (the precise
definition of $k_{in}$ is the coefficient appearing in Eq. (\ref{eq4}),
$V_{in}=k_{in}\Delta\rho gD^{2}/\eta_{in}$, but the value of $k_{in}$, 0.15,
used in the above is in fact the value of the coefficient $k_{in}^{\prime}$
appearing in the relation $V_{in}=(k_{in}^{\prime}\Delta\rho gD^{2}/\eta
_{in})^{\alpha}$ obtained when the data corresponding to the internal regime
in Fig. \ref{Fig2}(a) are numerically fitted by this relation with $\alpha$
determined to be not equal to one but close to 1.24, as mentioned in the first
paragraph in Experiment and Theory).\ In addition, the exponent in (\ref{eq8})
might also be slightly deviated from 3 as suggested in the paragraph just
above this one.

The crossover from the internal to external regime can explicitly be seen in
the data for $D=1.0$ mm (red data) in Fig. \ref{Fig2}(a). As $\eta_{in}$
decreases from the left-most data for $\nu_{in}=30000$ cS (red open diamonds)
to the data for $\nu_{in}=5000$ cS (red open inverse triangle), the velocity
is independent of $\nu_{in}$, which reveals that the three data on the
horizontal line are in the external regime. However, the data for $\nu
_{in}=1000$ cS and $\nu_{in}=500$ cS are on the straight dashed line with a
slope close to one, which confirms that these two data are in the internal
regime. Since the phase boundary expressed by Eq. (\ref{eq10}) is obtained
also by equating $V_{in}$ and $V_{ex}$ in Eqs. (\ref{eq4}) and (\ref{eq8}),
the crossover between the two regimes occurs in Fig. \ref{Fig2}(a) near at the
cross point between the horizontal line connecting the data on the external
regime for a given $D$ and the straight dashed line of a slope close to one
representing the internal regime.

The behavior of the data close to the crossover points are quite intriguing.
The data for $D=2.0$ mm and 3.0 mm at $\nu_{in}=1000$ cS (green filled square
and purple filled square) are located at the position close to the phase
boundary in Fig. \ref{Fig2}(c) (and the data have already been confirmed to be
in the internal regime in Fig. \ref{Fig2}(a): in this plot, these data points
are reasonably well on the dashed line).\ We have confirmed that, when these
two data are plotted in Fig. \ref{Fig2}(b), they are nearly on the straight
line of slope close to 3 in Fig. \ref{Fig2}(b). The two points can be
described by both Eqs. (\ref{eq4}) and (\ref{eq8}), which is reasonable
because they are nearly on the phase boundary. However, this is not always the
case. The data for $D=0.7$ mm and $\nu_{in}=5000$ cS (black filled inverse
triangle) and for $D=1.5$ mm and $\nu_{in}=3000$ cS (blue open triangle) are
also positioned close to the phase boundary in Fig. \ref{Fig2}(c). However,
the former is rather in the internal regime and the latter rather in the
external regime. This is in a sense logical because the blue open triangle is
rather away from the crossover point for $D=1.5$ mm in Fig. \ref{Fig2}(a) but
this is not the case for black filled inverse triangle. In general, how
quickly the crossover occurs seems to be a subtle problem.

\section*{Discussion}

\begin{figure}[ptb]
\begin{center}
\includegraphics[width=\linewidth]{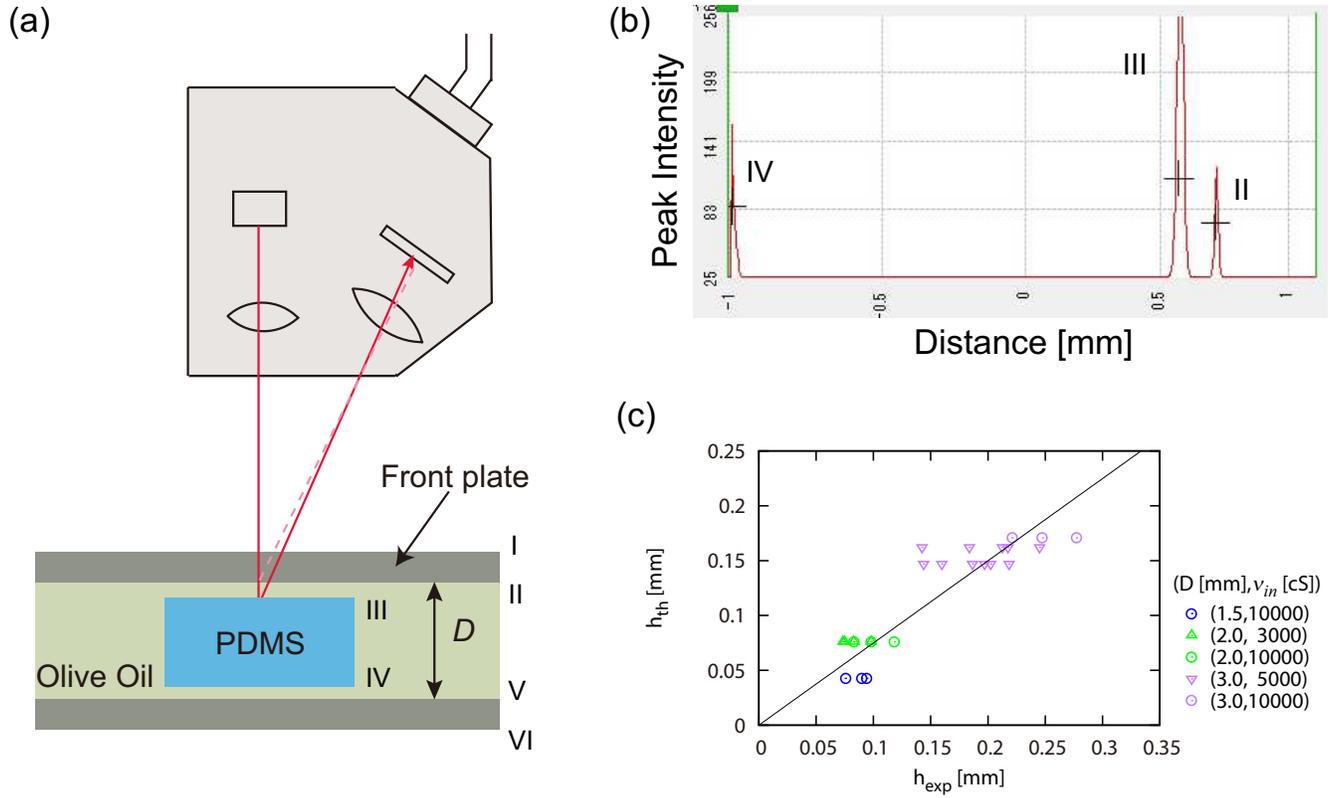}
\end{center}
\caption{(a) Setup for the thickness measurement. (b) Example of the screen
shot of the three peaks (the left to the right) originating from interface IV,
III and II (see the text for the details). The distance between IV and III is
obtained by multiplying $n_{PDMS}$ to the distance, whereas that between III
and II is obtained via $n_{olive}$ instead. (c) Plot of the experimentally
obtained value of the thin film thickness $h_{exp}$ vs. the theoretical
estimation $h_{th}$. The theoretical value $h_{th}$ is obtained by Eq.
(\ref{eq9}) with the coefficient $k_{2}=k_{LL}^{1/2}k_{1}$ with $k_{1}=0.15$
and $k_{LL}=0.94$.}%
\label{Fig3}%
\end{figure}

The direct measurement of the thickness $h$ supports the above analysis. We
used a laser distance sensor\ (ZS-HLDS2+ZS-HLDC11+Smart Monitor Zero Pro.,
Omron), as illustrated in Fig. \ref{Fig3}(a). The measurement is extremely
delicate and difficult, because we have six reflective planes I to VI with
significantly different strengths of reflection where the target two
reflections II and III are the smallest and the second smallest among them
(see Fig. \ref{Fig3}(b)). The six surfaces are the front and back surfaces of
the front cell plate (interface I and II), the front and back interfaces
between olive oil and the PDMS drop (interface III and IV), and the front and
back surfaces of the back cell plate (interface V and VI). To determine $h$,
we need to detect reflection from interface II and III, where the
reflection\ from II is small compared with that of III (see Fig.
\ref{Fig3}(b)) and significantly small compared with that of I, because the
refractive index of olive oil is $n_{olive}=1.47$, that of acrylic plate is
$n_{acr}=1.491$, that of PDMS oil is $n_{PDMS}=1.403$ and that of air is
$n_{air}=1$. Furthermore, the object (the descending drop) is moving. In spite
of these experimental difficulties, we obtained a reasonably good correlation
between the measured thickness and the experimentally\ obtained value as shown
in Fig. \ref{Fig3}(c), by virtue of various efforts (for example, in the
screen shot Fig. \ref{Fig3}(b), the two target peaks are intensionally
positioned off-center because the precision of measurement becomes the maximum
when the reflection angle is the largest). Here, the slope of the line
obtained by a numerical fitting is $0.749\pm0.027$ (the slope here is not the
exponent but the coefficient for the linear relationship), the order of
magnitude of which is consistent with the scaling argument.

Exceptional data mentioned above reveal an intriguing phenomenon. In Fig.
\ref{Fig2}(a), the data for $D=1$ mm and for $\nu_{in}=10000$ cS are
represented by two different marks, the red filled circle and the red open
circle, with the former described by the internal regime and the latter by the
external regime. The data for $D=1$ mm and for $\nu_{in}=5000$ cS are also
categorized into two filled and open symbols. The experimental difference in
acquiring these two different types (filled and open symbols) of data obtained
for identical drop viscosity and cell spacing\ is that, when the drop goes
down on the same path multiple times in the same cell, the first drop is in
the external regime (open marks) whereas the drop going down after the first
one is always in the internal regime (filled marks). This apparently
mysterious effect is quite reproducible and is understood by considering a
possibility of mixing of olive oil and PDMS at the surface of the drops. For
the first drop, such a mixing effect is negligible and the drop is governed by
the dynamics of the external regime. However, after the first one, because of
the mixing effect, the viscosity of the thin film surrounding the drop
increases (because $\nu_{in}\gg\nu_{ex}$)\ so that making a velocity gradient
in the external thin film is no longer favored in terms of energy and instead
the velocity gradient inside the drop is favored to realize the dynamics in
the internal regime. Because of this reason, the red filled circles\ and the
red filled inverse triangles\ are not shown in the phase diagram given in Fig.
\ref{Fig2}(c). This seemingly mysterious behavior tends to be suppressed if
the viscosity is too small (because the "external" viscosity does not get
sufficiently viscous), or too large (because the mixing is not sufficiently
effective). This is why we observed this phenomenon only for the two values of viscosity.

The present study suggests that Stokes' drag friction $F=6\pi\eta_{ex}VR$ for
a solid sphere of radius $R$ surrounded by a viscous liquid of viscosity
$\eta_{ex}$ is replaced in the external regime of the Hele-Shaw cell geometry
by
\begin{equation}
F_{ex}\simeq\eta_{ex}VR_{T}R_{L}/h\simeq\eta_{ex}Ca^{-2/3}VR_{T}R_{L}%
/\kappa^{-1}.\label{eq11}%
\end{equation}
This expression possesses a nonlinear dependence on the velocity $V$ due to
the extra $V$ dependence contained in the capillary number $Ca$. This is
strikingly different from the two other expressions for the drag force:
$F_{in}\simeq\eta_{in}VR_{T}R_{L}/D$ and $F_{bubble}\simeq\eta_{ex}VR_{T}%
^{2}/D$, which are both linear in velocity; The former corresponds to the
internal regime in the present study, whereas the latter corresponds to the
case in which the dominant dissipation is the one associated with the velocity
gradient $V/D$ in the surrounding external liquid \cite{EriSoftMat2011}. The
viscous friction forces including the nonlinear friction in Eq. (\ref{eq11})
are relevant to the dynamics of emulsion, foam, antifoam and soft gels
\cite{Sylvie,AnnLaureSylvieSM2009,DominiqueMicrogravity2015}, in particular,
nonlinear rheology of such systems
\cite{DenkovSoftMat2009,DurianPRL10,CloitreNP2011}.

We intentionally used several times the phrase, \textquotedblleft the order of
magnitude of which is consistent with the scaling argument,\textquotedblright%
\ which may be vague compared with an expression like, \textquotedblleft being
of order one further supports the scaling argument.\textquotedblright\ The
reason we used the seemingly vague expression is that whether a coefficient
for a scaling law is of the order of one or not is in fact a subtle issue.
Depending on the problem or on the definition of the coefficient, the orders
of magnitude can be fairly larger or smaller than one. An example of such a
case can be given by exploiting the relation, $k_{3}=k_{1}^{3}/k_{in}$, given
above: The three coefficients, $k_{1},k_{3}$, and $k_{in}$ are all
coefficients for some scaling laws so that, for example, $k_{1}$ and $k_{in}$
can be 5 and 1, respectively, but this example implies $k_{3}$ is much larger
than one ($k_{3}=5^{3}$).\ 

In the present study, the consistency of the whole scaling arguments is
checked in several ways, which clearly deepens our physical understanding. For
example, a new scaling regime is demonstrated through a clear data collapse
(Fig. \ref{Fig2}(b)), and the crossover of this regime to another is shown
(Fig. \ref{Fig2}(a)), which is completed by the phase diagram (Fig.
\ref{Fig2}(c)) and a separate measurements on thin-film thickness (Fig.
\ref{Fig3}(c)). In addition, data arrangements in the crossover diagram (Fig.
\ref{Fig2}(a)) are interpreted from various viewpoints, confirming the
consistency of the arguments.

\section*{Conclusion}

In summary, we show in Fig. \ref{Fig2}(b) the existence of a novel scaling
regime for the descending velocity of a drop surrounded by thin external fluid
in the Hele-Shaw cell, in which regime the viscous dissipation in the thin
film is essential. This regime corresponds to a nonlinear form of viscous drag
friction. In this regime, the thickness of the film is determined by the law
of LLD, as directly confirmed in Fig. \ref{Fig3}(c). The crossover between
this regime and another regime in which the viscous dissipation in the
internal side of the drop governs the dynamics is shown in Fig. \ref{Fig2}(a).
The phase boundary between the two regimes are given in Fig. \ref{Fig2}(c).

There are some other scaling regimes for the viscous drag friction in the
Hele-Shaw cell geometry with the existence of thin films surrounding a fluid
drop. For example, the dissipation associated with the velocity gradient $V/D$
in the internal drop liquid has been revealed to be important for a rising
bubble in the Hele-Shaw cell \cite{EriSoftMat2011}. The dissipation associated
with the dynamic meniscus (in the context of LLD theory
\cite{LandauLevich,Derjaguin1943,CapilaryText}) formed in the external
thin-film has been found to be important in a non Hele-Shaw cell geometry
\cite{PascalEPL2002}. In addition, the present external regime will give
another scaling law if the capillary length $\kappa^{-1}$ is, unlike in the
present study, larger than the cell thickness $D$.

Confirmation of such other regimes for viscous drag friction in the Hele-Shaw
cell geometry, as well as crossovers among various scaling regimes would be
explored in future studies. The simple friction laws for confined fluid drops
and the crossover between them revealed in the present study (and in future
studies) are relevant to fundamental issues including rheology of foam and
emulsion, as well as applications such as in microfluidics.

\section*{Methods}

The density of PDMS oil $\rho_{in}$ slightly depends on viscosity: (1) 970
kg/m$^{3}$ for the kinematic viscosities $\nu_{in}=500,1000,$ and $3000$ cS
(SN-4, SN-5, and SN-6, As One). (2) 975 kg/m$^{3}$ for $\nu_{in}=5000$ and
$10000$ cS (SN-7 and SN-8, As One). (3) 976 kg/m$^{3}$ for $\nu_{in}=30000$ cS
(KF-96H, ShinEtsu).

The cell thickness $D$ is controlled by spaces, and is directly measured using
the laser distance sensor\ (ZS-HLDS5, Omron) for most of the cells. In all the
figures of the present study, for simplicity, the cell thickness $D$ is
represented by an approximate value, which is slightly different from measured
values. For some of the data the measurement of $D$ was not performed and in
such a case an approximate value of $D$ is used, instead of measured values,
to plot the data points, which does not cause serious difficulties in
analyzing and interpreting the data. This is because the difference between
the $D$ value used for labeling and the measured value of the cell thickness
is rather small.

The interfacial tension between PDMS and olive oil was measured by using
pendant drop tensiometry. It is recently discussed that measured values for
pendant drops are dependent on Bond number and Worthington number, with both
scaling with $B=\Delta\rho gR_{0}^{2}/\gamma$ ($R_{0}$: the drop radius at the
apex of the pendant drop) when the drop size is of the same order of magnitude
as the needle diameter, and that the measured value approach the correct value
as $B$ approaches one \cite{PendantDrop2015} (one could expect that the
experimental precision will be optimized when the drop is most "swelled," that
is, when the droplet is on the verge of detaching off from the needle tip due
to gravity, that is, when $B=1$).\ We measured the value of tension as a
function of $B$ by using the software, OpenDrop, developed by Michael Neeson,
Joe Berry and Rico Tabor. We extrapolated the data thus obtained to the value
at $B=1$ to have a pragmatic\ value, $\gamma=0.78$ mN/m, because it was
experimentally difficult to approach $B=1$. This is possibly because the
tension is significantly small, which might lead to an extra error in the measurement.

Even though the measurement of the interfacial tension contains an extra error
and our analysis numerically depends on the measured value, this does not
bring any uncertainties in the present arguments at the level of scaling laws.
We explain this by an example. Introducing the experimentally measured value
of surface tension $\gamma_{m}$, we define a numerical coefficient $\beta$ as
$\gamma=\beta^{2}\gamma_{m}$ and the corresponding capillary length
$\kappa^{-1}=\beta\kappa_{m}^{-1}$. With these "measured" quantities, Eq.
(\ref{eq8}) can be expressed as $\eta_{ex}V_{ex}/\gamma_{m}=k_{1,m}%
^{3}(D/\kappa_{m}^{-1})^{3}$ with $k_{1,m}^{3}=k_{1}^{3}/\beta$. By noting
that the values of the interfacial tension and capillary number used in Fig.
\ref{Fig2}(c) that experimentally confirms the relation Eq. (\ref{eq8}) are in
fact not $\gamma$ and $\kappa^{-1}$ but $\gamma_{m}$ and $\kappa_{m}^{-1}$,
respectively, the coefficient we determined from Fig. \ref{Fig2}(c) is in fact
not $k_{1}$ but $k_{1,m}$. However, since Eq. (\ref{eq10}) can be expressed as
$\eta_{ex}/\eta_{in}=k_{3,m}D/\kappa_{m}^{-1}$ with $k_{3,m}=k_{3}/\beta$, the
phase boundary line $\eta_{ex}/\eta_{in}=k_{3,m}D/\kappa_{m}^{-1}$ on the
$(\eta_{ex}/\eta_{in},D/\kappa_{m}^{-1})$ space and the line $\eta_{ex}%
/\eta_{in}=k_{3}D/\kappa^{-1}$ on the $(\eta_{ex}/\eta_{in},D/\kappa^{-1})$
space have the same physical meaning. From these reasons, a special care is
needed when one compares the numerical coefficient obtained experimentally in
the present study with more sophisticated experiments or calculations.


\section*{Acknowledgements}

This research was partly supported by Grant-in-Aid for Scientific Research (A)
(No. 24244066) of JSPS, Japan, and by ImPACT Program of Council for Science,
Technology and Innovation (Cabinet Office, Government of Japan).

\section*{Author contributions statement}

K.O. and N.K. conceived the experiment, and N.K. collected initial data while
M.Y. conducted the ensuing experiments. M.Y. and K.O. analyzed the results,
M.Y. and K.O. prepared the figures and graphs, K.O. wrote the manuscript. All
authors reviewed the manuscript.

\section*{Additional information}

Competing financial interests: The authors declare no competing financial interests.

\clearpage

\end{document}